\documentstyle[aps,multicol,epsfig,prl,floats]{revtex}

\begin{document}
\draft
\twocolumn[\hsize\textwidth\columnwidth\hsize\csname @twocolumnfalse\endcsname
\title{Metallic nonsuperconducting phase and d--wave superconductivity\\
in Zn--substituted La$_{1.85}$Sr$_{0.15}$CuO$_4$}
\author{K. Karpi\'{n}ska $^{1}$, Marta Z. Cieplak $^{1,2}$, S. Guha $^{2}$, A.
Malinowski $^{1}$, T. Sko\'{s}kiewicz $^{1}$,\\ W. Plesiewicz $^{1}$, M.
Berkowski $^{1}$, B. Boyce$^{3}$, Thomas R. Lemberger$^{3}$, and P.
Lindenfeld $^{2}$}
\address{$^1$ Institute of Physics, Polish Academy of Sciences, 02 668
Warsaw, Poland\\
$^{2}$ Department of Physics and Astronomy, Rutgers University, Piscataway,
NJ 08854--8019, USA\\
$^{3}$ Department of Physics, Ohio State University, Columbus, OH
43210--1106, USA}
\maketitle

\begin{abstract}
Measurements of the resistivity, magnetoresistance and penetration depth
were made on films of La$_{1.85}$Sr$_{0.15}$CuO$_{4}$, with up to 12
at.\% of Zn substituted for the Cu. The results show that the quadratic
temperature dependence of the inverse square of the penetration depth,
indicative of d--wave superconductivity, is not affected by doping. The
suppression of superconductivity leads to a metallic nonsuperconducting
phase, as expected for a pairing mechanism related to spin fluctuations. The
metal--insulator transition occurs in the vicinity of $k_F l \approx 1$, and
appears to be disorder-driven, with the carrier concentration unaffected by
doping.
\end{abstract}


]

Although there is strong evidence for d--wave symmetry of the
order parameter in high--T$_c$ superconductors \cite{annett1}, earlier experiments
do not distinguish between mechanisms that lead to pure
$d_{x^{2}-y^{2}}$ symmetry \cite{annett2}, and others that allow an 
admixture of s--wave pairing \cite{tolpy1}.

In this letter we describe the suppression of superconductivity by
disorder, with the conclusion that pure d--wave symmetry continues
until the superconductivity disappears. The experiment is based on the
fact that disorder strongly suppresses d--wave pairing, and may
therefore lead to a transition from the superconducting state to a
normal--metal state. Any s--wave pairing would be less strongly
affected, so that in its presence superconductivity would be expected
to persist  until, with greater disorder, it is destroyed at the
metal--insulator transition.

Studies of the $T_c$--suppression in
electron--irradiated YBa$_2$Cu$_3$O$_{7-\delta}$ (YBCO) \cite{tolpy1}, or in
Zn--doped YBCO and La$_{2-x}$Sr$_{x}$CuO$_4$ (LSCO) \cite{fuku} did not
address the question of the nature of the nonsuperconducting phase. Our
previous study showed the existence of a metallic nonsuperconducting
phase in  LSCO with variouys impurities \cite{mi}, but was subject to
criticism because it was done on polycrystalline, ceramic specimens.

We studied a series of
single--crystalline La$_{1.85}$Sr$_{0.15}$Cu$_{1-y}$Zn$_{y}$O$_{4}$ films,
with zinc content, $y$, from 0 to 0.12, and complete suppression of
superconductivity for $y>$ 0.055. We find that with increasing $y$ the
transition from the superconducting state is to a metallic state, with
the carrier concentration unaffected by the addition of the zinc. This
is in  contrast with the carrier--driven transition that is
observed with a change in the strontium content. We also measured the
superconducting penetration depth ($\lambda $), and find that it
remains proportional to $T^{2}$ when $y$ is 
increased, suggesting that there is no s--wave component.
As $y$ increases to 0.12, the 
metal--insulator transition is approched in the vicinity of 
$k_F\ell = 1$, where $k_F$ is the Fermi wave vector and $\ell$ is the
electronic mean free path, suggesting that the transition is disorder--driven. 

The $c$--axis oriented films, with thicknesses between 5000 and 9000 {\AA }
, were grown by pulsed laser deposition on LaSrAlO$_{4}$ substrates. The
films were patterned by photolithography and wires were attached with indium
to evaporated silver pads. Standard six--probe geometry was used to measure
the Hall voltage and the magnetoresistance. The specimens were mounted in a
dilution refrigerator, and cooled to 20 mK without a magnetic field, and to
45 mK in the presence of a field. The magnetoresistance was measured with
low--frequency ac, in magnetic fields up to 8.5 T, in the longitudinal
(field parallel to the $ab$--plane) and transverse (field perpendicular to
the $ab$--plane and to the current) configurations. A second set of
specimens was prepared for penetration depth measurements. $\lambda (T)$ was
obtained from the mutual inductance of two coaxial coils fixed on opposite
sides of the superconducting film \cite{turneaure}.

The zinc content in the films was checked to confirm that it is the same as
that of the targets \cite{ciep}. The films have some substrate--induced
strain, and varying amounts of oxygen vacancies so that they had 
a range of resistivities at each value of y \cite{ciep}.
For each y a group of 6 to 10 films was made, and it was
possible to select films with residual resistivities within 30\% of
those for bulk single crystals \cite{fuku}. In these selected
films superconductivity persists to $y_c$ = 0.055, while for larger
resistivities $T_c$ vanishes earlier. In ceramic specimens $y_c$ = 0.03
\cite{xiao}. 

\begin{figure}[ht]
\epsfig{file=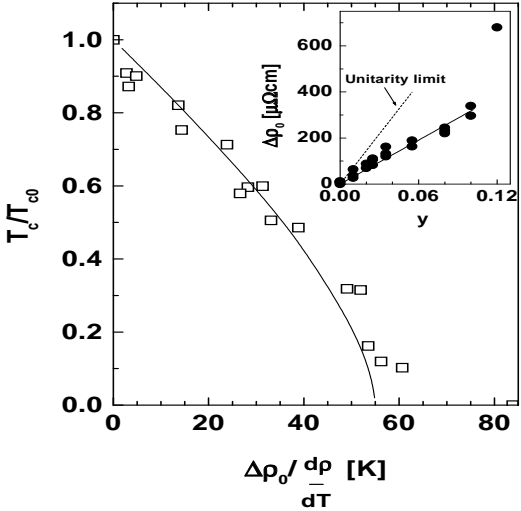, height=0.4\textwidth, width=0.4\textwidth}
\caption{The normalized critical temperature, $T_c /T_{c0}$, as a function of
$\Delta \rho_0 /{\frac{d\rho }{dT}}$ for a set of Zn--doped
LSCO films, with $0 < y < 0.055$. 
The solid line shows the best fit to the AG formula with $\lambda_{TR}$ =
0.18.
Inset: The dependence of $\Delta \rho_0$ on $y$. The solid line is a fit to
the
experimental points for $0 < y <0.10$. The dashed line indicates the
unitarity limit.}
\label{Fig.1}
\end{figure}

The increase in the residual resistivity, $\Delta \rho _{0}$,
as a function of $y$ is shown in the inset in Fig.~1.
The residual resistivity is determined by extrapolation to zero temperature
of the linear high--temperature resistivity, and $\Delta \rho _{0}$ is
calculated with respect to a zinc--free film with $\rho _{0}$=36.5$\mu
\Omega cm$ and $T_{c0}$=35.2K. It is seen that $\Delta \rho _{0}$ increases
linearly with $y$ at a rate of 3.3 $\mu \Omega cm$ per at.\% of Zn until $y$
reaches 0.1. For larger concentrations the rate increases rapidly, signaling
the approach to the metal--insulator transition. The resistivity
in a two--dimensional system from s--wave impurity scattering follows
the formula $\Delta \rho _{0}=4(\hbar/e^{2})(y/n)~d~sin^{2}{\delta _{0}}$,
where $n$ is the carrier concentration, 
$d$ is the distance between the CuO$_{2}$ planes (6.5 {\AA} 
in LSCO), and $\delta _{0}$ is a phase shift \cite{chien}. The dashed line
shows the unitarity
(maximal) limit corresponding to $\delta _{0}=\pi /2$. We use $n = 0.15$,
as the carrier concentration is shown by Hall--effect measurements 
to be almost independent of $y$ \cite{art}. It is seen that the 
scattering is about half
as effective as in the unitarity limit. This result
is close to the result for Zn--doped YBCO single crystals \cite{chien}, but
differs from that reported for single crystals of
Zn--doped LSCO where the scattering was claimed to exceed the unitarity
limit \cite{fuku}. In fact, the discrepancy seems to be primarily the
result of a difference in the method of calculating the residual
resistivities \cite{comment}.

The scattering rate for nonmagnetic impurities, ${1}/{\tau _{IMP}}$, is
related to the residual resistivity, $\Delta \rho _{0}$, by the formula $%
1/\tau _{IMP}=2\pi \lambda _{TR}k_{B}\Delta \rho _{0}/\hbar {\frac{d\rho }{dT%
}}$, where $\lambda _{TR}$ is an electron--boson transport coupling
constant, and the value of $\frac{d\rho }{dT}$ is from the temperature range
200K to 300K where the resistivity is boson--mediated. We display the $T_{c}$%
--suppression by plotting the ratio $\Delta \rho _{0}/{\frac{d\rho }{dT}}$,
since the errors related to uncertainties of size and homogeneity of the
specimens cancel in this ratio \cite{tolpy1}. This is shown in Fig.~1, where 
$T_{c}$ is normalized to $T_{c0}$ = 35.2 K. The solid
line shows the best fit to the Abrikosov--Gorkov (AG) formula
\cite{abrik}, $\ln \frac{%
T_{c0}}{T_{c}}=\Psi \left( \frac{1}{2}+\frac{\hbar }{4\pi \tau k_{B}T_{c}}%
\right) -\Psi \left( \frac{1}{2}\right) $ (where $\Psi $ is the digamma
function), with the pair--breaking scattering rate $1/\tau $ equal to $%
1/\tau _{IMP}$ \cite{millis,radtke} and the fitting parameter $\lambda _{TR}$
equal to 0.18. There is a slight deviation of the theoretical curve from the
experimental points on Fig.~1 for $T_{c}/T_{c0}<0.25$. The value of $\lambda
_{TR}$ of 0.18 (indicating the weak--coupling limit) is a factor of two
smaller than the value we estimate from high--temperature resistivity data,
for $\frac{d\rho }{dT}$ equal to the average of the experimental values, 2.5$%
\mu \Omega $ cm/K, and $\hbar \omega _{P}=0.8eV$ \cite{uchida},\cite{gurvitch}.

\begin{figure}[ht]
\epsfig{file=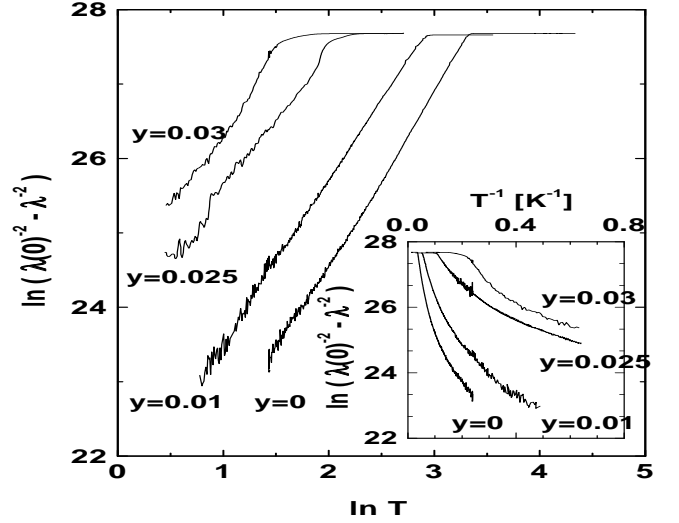, height=0.4\textwidth, width=0.5\textwidth}
\caption{Penetration depth results: log--log graph of ${\lambda (0)}^{-2} -
\lambda^{-2}$ versus temperature, for films with $y$ = 0, 0.01, 0.025,
and 0.03 and $\lambda (0)$ = 9800 $\AA$, 10000 $\AA$, 16000 $\AA$,
19300 $\AA$, respectively. Inset: $\ln{({\lambda (0)}^{-2} - \lambda^{-2})}$
versus $1/T$ for the same films.}
\label{Fig.2}
\end{figure}

Both the deviation of the curve for small $T_{c}$ and the reduced value of $%
\lambda _{TR}$ may be related to effects which are not included in the AG
formula, such as the anisotropy of the impurity scattering (which leads to a
difference between $1/\tau $ and $1/\tau _{IMP}$ \cite{haran}) and the
spatial variation of the order parameter \cite{franz,zhito}. The
dependence of $T_{c}/T_{c0}$ on $\Delta \rho _{0}$ by itself does not allow
us to distinguish between these effects. The deviations could also be
caused by the presence of an s--wave component of the superconducting order
parameter, and this possibility is examined below.

Fig.~2 shows the temperature dependence of the penetration depth for four
films with values of $y$ between 0 and 0.03, and values of $T_{c}$ from 28 K
to 6.2 K. The specimen with the lowest $T_{c}$ has $T_{c}/T_{c0}$ equal to
0.18 which is in the regime where the AG formula begins to deviate from the
experimental data (see Fig.~1). For d--wave symmetry of the order parameter,
disorder leads to a quadratic temperature dependence $\lambda
^{-2}(T)-\lambda ^{-2}(0)\propto T^{2}$ for $T\ll T_{C}$, while exponential
behavior, $\lambda ^{-2}(T)-\lambda ^{-2}(0)\sim \exp {(-\Delta }_{{%
\min }}{/kT)}$, is expected for $T\ll \Delta _{\min }$ in an s--wave
superconductor \cite{annet3,hirsch}. Here the magnitude of the minimum of
the energy gap, $\Delta _{\min }$, is expected to increase with
disorder for anisotropic s--wave pairing \cite{andreone,borko}. The
quadratic temperature dependence is well documented for zinc--free LSCO
films with different amounts of intrinsic disorder \cite{paget}. It is
evident from Fig.~2 that the quadratic dependence gives a much better
description of the data for all values of $y$. An attempt to fit a straight
line to the low--temperature portion of the curves, as shown in the inset,
leads to energy gaps decreasing with increasing $y$, inconsistent with the
expectations for anisotropic s--wave symmetry. A similarly negligible effect
of zinc doping on the $T$--dependence of $\lambda $ has also been reported
for YBCO \cite{ulm}. We note that nonmagnetic disorder should produce a
rapid decrease of the zero--temperature value of the superfluid density, $%
n_{S}\left( 0\right) \sim \lambda (0)^{-2}$, together with a steep decrease
of $T_{c}$ \cite{sun}. Our results show that $n_{S}(0)$ decreases by a
factor of 1.3 per percent of impurity, which is slightly slower than the
rate reported for YBCO \cite{ulm} but close to a theoretical
predictions for a d--wave superconductor \cite{sun}.

\begin{figure}[ht]
\epsfig{file=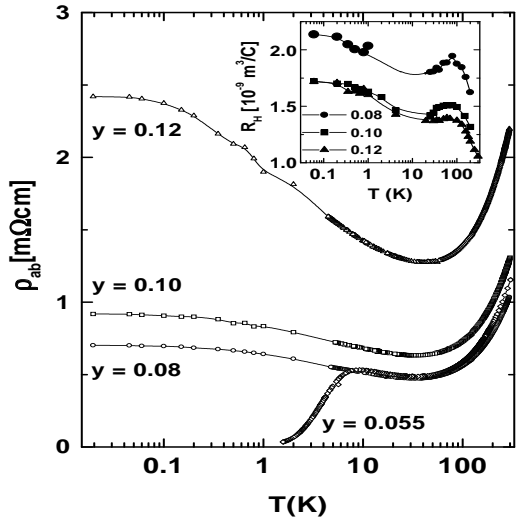, height=0.4\textwidth, width=0.4\textwidth}
\caption{The temperature dependence of the resistivity for four films with
values of
$y$ of 0.055, 0.08, 0.1, and 0.12. Inset: Hall coefficient for
films with $y$ = 0.08, 0.1, and 0.12 as a function of temperature.
All lines are guides to the eye.}
\label{Fig.3}
\end{figure}

Fig.~3 shows the temperature dependence of the resistivity down to 20 mK for
several films with large amounts of zinc. The film with $y$ = 0.055 is still
superconducting, with a transition whose midpoint is at 3.5 K. The
films with larger values of $y$ (0.08, 0.10, and 0.12), are
nonsuperconducting, and their resistivity increases approximately as $%
\rho _{ab}\sim \ln {(1/T)}$ as $T$ is lowered, but below about 300 mK the
increase slows down and the resistivity is clearly finite in the $T=0$
limit. This behavior is markedly different from the evolution of the $ab$%
--plane resistivity with strontium content, where nonsuperconducting
specimens exhibit hopping conductivity indicative of insulating behavior in
the $T=0$ limit.

The inset to Fig.~3 shows the temperature dependence of the Hall
coefficient, $R_H$, for three nonsuperconducting films. A slow increase
of $R_H$ followed by saturation below 300 mK is seen in the low--temperature
region. The magnitude of $R_H$ is close to that observed for $y = 0$ 
\cite{hwang}, indicating that the specimens remain metallic up to the
highest doping level, $y = 0.12$, without any change in the carrier
concentration.

\begin{figure}[ht]
\epsfig{file=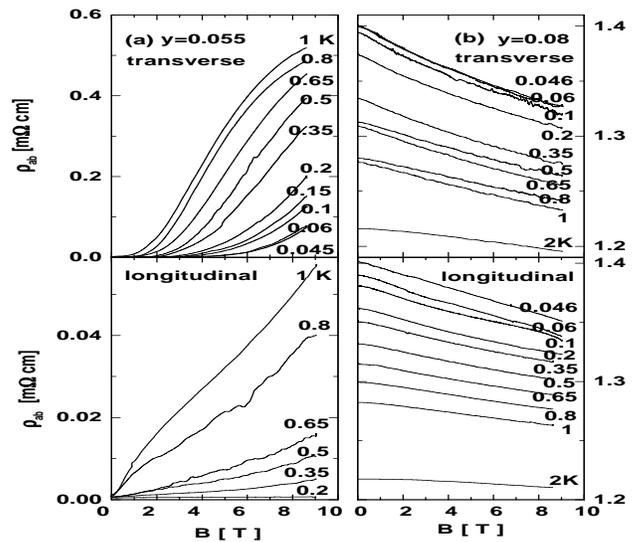, height=0.45\textwidth, width=0.6\textwidth}
\caption{Transverse and longitudinal magnetoresistance for films with $y$ =
0.055
(a), and $y$ = 0.08 (b) versus magnetic field. The labels show the
temperatures at
which the measurements were made.}
\label{Fig.4}
\end{figure}

The saturation of the resistivity shown in Fig.~3 could signal
some remanence of the superconducting phase. We examined this possibility by
magnetoresistance measurements. Fig.~4 shows the resistivity as a function
of magnetic field, applied in the longitudinal and transverse
configurations, for two films, with $y$ equal to 0.055 and 0.08, which are
superconducting and nonsuperconducting, respectively, in the absence of the
field. The curves are for constant temperatures from 2 K to 45 mK. The
magnetoresistance is positive for the film with $y$ = 0.055 for both field
configurations. This is similar to the magnetoresistance in LSCO without
zinc \cite{karp} and is clearly caused by the suppression of
superconductivity by the magnetic field. On the other hand the
magnetoresistance of the film with $y$ = 0.08 is negative for both
configurations, with the magnitude of the transverse magnetoresistance about
twice as large as the longitudinal. If we attribute the longitudinal
magnetoresistance entirely to spin--related isotropic scattering, the
difference between the transverse and longitudinal magnetoresistance gives
the orbital part, which is then also negative. This result demonstrates that
superconducting fluctuations are absent in the specimen with $y$=0.08 and
that the metallic nonsuperconducting phase is uniform without any
macroscopic superconducting inclusions.

The values of the conductivity at $T$ = 20 mK, $\sigma _0$, for three
metallic films with $y$ = 0.08, 0.10, and 0.12, are equal to 1430,
1098, and 436 $(\Omega cm)^{-1}$, respectively. Using a linear 
extrapolation to $\sigma _0$ = 0 we estimate
that the metal--insulator transition occurs at $y_{MI}\approx $ 0.14.
We also estimate the magnitudes of $k_F\ell$ for these specimens as equal to
2.4, 1.8, and 0.7, respectively from the relation for a 2D free--electron 
system, $k_F\ell = h d \sigma _0 /e^2$. We see that
the metal-insulator transition occurs in the vicinity of $k_F\ell = 1$,
as expected for disorder--induced localization 
\cite{lr} in contrast to suggestions that the
superconductor--insulator transition in this system occurs at
$h/4e^2$ = 6.5 k$\Omega$ \cite{fuku}. The value of $y_{MI}$ is
remarkably small compared to the fraction of the nonmetallic
constituent required for the metal--insulator transition in amorphous
systems \cite{beki}, showing that the Zn creates extremely
effective localization centers in the CuO$_2$ plane
(see, e. g., Ref. \cite{Tallon})
and that in addition to the effect of disorder,
the scattering by Zn--impurities is enhanced by electron-electron
interactions in this strongly correlated system.

In related work we have shown that with increasing $y$ the high--temperature
susceptibility of the ceramic targets evolves toward the Curie--Weiss
relation, $\chi = C/(T + \Theta )$, with $\Theta$ reaching about 40K
at $y$ = 0.14 \cite{malin}. This result indicates that large Zn--doping, 
while removing Cu--spins, restores some 
antiferromagnetic ordering in LSCO. It has been suggested before on 
the basis of neutron scattering experiments that substitution of a 
small amount of zinc in LSCO ($y$ = 0.012) may stabilize a short—-range--order 
spin--density--wave state \cite{hirota}, and our result appears to be 
consistent with this suggestion. Related magnetic effects are presumably 
responsible for the large negative magnetoresistance and for the 
saturation of the resistivity at low temperatures.

The experiments described here indicate that the
suppression of 
superconductivity by Zn--doping in LSCO proceeds without a change of the
symmetry of the order parameter, up to the point
where superconductivity disappears and the normal metallic phase is reached.
This result favors models that predict pure d--wave symmetry, as for
example, the spin--fluctuation exchange model. With further doping the
metal--insulator transition occurs in the vicinity of $k_F\ell = 1$, 
showing that it is disorder--driven.

We would like to thank J. Annett and A. Millis for helpful discussions, and
Elisabeth Ditchek for help with the specimen preparation. This work was
supported by the Polish Committee for Scientific Research, KBN, under grants
2P03B 09414 and 2P03B 10714, and by the Naval Research Laboratory. Work at
OSU was supported by DOE grant No. DE-FG02-90ER45427 through the Midwest
Superconductivity Consortium.


\begin{references}
\bibitem{comment}  Using the definition employed in this work as well
as in Ref.  
\cite{chien}, the values of $\rho _{0}$ that we calculate from the data
of Fig.~1 of Ref. \cite{fuku} are about half of those reported in
Fig.~4 of 
that paper, so that they are comparable to our results.
\end{references}
\end{document}